\documentclass[letterpaper,aps,pre,twocolumn,showpacs,preprintnumbers,
superscriptaddress]{revtex4}

\usepackage{graphicx}
\usepackage{epsf} 
\usepackage{dcolumn}
\usepackage{bm}
\usepackage{amssymb}
\usepackage{amsmath}
\usepackage{amsbsy}

\usepackage{graphicx,color,ifthen}
\usepackage{pstricks,pst-node,pst-text,pst-3d,color}

\newcommand{\Ge}{\Gamma_{ex}}
\newcommand{\Ga}{\Gamma_{ad}}

\begin{document}

\title{Short range stationary patterns and long range disorder in an evolution
equation for one-dimensional interfaces}

\author{Javier Mu{\~n}oz-Garc\'{i}a}
\affiliation{Departamento de Matem\'aticas and Grupo Interdisciplinar de
Sistemas Complejos (GISC), Universidad Carlos III de Madrid, Avenida de la
Universidad 30, E-28911 Legan\'es, Spain}

\author{Rodolfo Cuerno}
\affiliation{Departamento de Matem\'aticas and Grupo Interdisciplinar de
Sistemas Complejos (GISC), Universidad Carlos III de Madrid, Avenida de la
Universidad 30, E-28911 Legan\'es, Spain}

\author{Mario Castro}
\affiliation{GISC and Grupo de Din\'amica No Lineal (DNL), Escuela T\'ec.\
Sup.\ de Ingenier{\'\i}a (ICAI), \\ Universidad Pontificia Comillas, E-28015
Madrid, Spain}

\date{\today}

\begin{abstract}
A novel local evolution equation for one-dimensional interfaces is
derived in the context of erosion by ion beam sputtering. We
present numerical simulations of this equation which show
interrupted coarsening in which an ordered cell pattern develops
with constant wavelength and amplitude at intermediate distances, while
the profile is disordered and rough at larger distances. Moreover,
for a wide range of parameters the lateral extent of ordered
domains ranges up to tens of cells. This behavior is new in the
context of dynamics of surfaces or interfaces with morphological
instabilities. We also provide analytical estimates for the
stationary pattern wavelength and mean growth velocity.
\end{abstract}

\maketitle

Pattern formation is ubiquitous in nature and one of the most
fascinating features of nonequilibrium systems \cite{cross:1993}.
Typical examples can be found in many startlingly similar
interfaces which emerge in very different processes, such as
growth of amorphous \cite{barabasi} and epitaxial thin films
\cite{politi00}, or erosion by ion beam sputtering (IBS)
\cite{valbusa:2002}. These pattern forming surfaces can be
classified into different categories according to the stationary
or time-dependent
 (coarsening) behavior of the typical pattern length scale $l$.
 Actually, a large effort has been
devoted recently to describe these systems through height
equations \cite{note1}, since these provide compact and efficient
analytical/numerical
 descriptions that successfully describe global morphological properties such as
kinetic roughening, and surface pattern formation and coarsening.
 Specifically, in order to assess the
 predictive power of ($1d$) height equations, it would be important to produce
criteria for the presence or absence of coarsening. In Ref.\
\cite{politi} up to four different scenarios have been proposed
depending on the behavior of $l$ as a function of the pattern
amplitude $A$. The conclusion is that coarsening stops ({\em
interrupted coarsening}) if the function $l(A)$ attains a maximum,
after which the amplitude increases indefinitely with time. A
similar conclusion was reached at by Krug in Ref.\
\cite{krug:2001}, where it is actually conjectured that no ($1d$)
{\em local} height equation can ``describe the emergence and
evolution of patterns with constant wavelength {\em and}
amplitude''.
Hence, a $1d$ counterexample of an interface equation \cite{note1} leading to a
stationary pattern with constant wavelength and amplitude would be interesting
to
improve our understanding of interrupted coarsening and, specifically, of
the type of nonlinearities that induce it \cite{castro06}.

Recently, a continuum 2d model has been introduced which describes interesting
(sub)micrometric features of surfaces eroded by IBS \cite{castro05, munoz:2006}.
The resulting interface equation was studied numerically for moderate system
sizes
suggesting the occurrence of a stationary ordered pattern and interrupted
coarsening
\cite{note2}. Here, we derive the $1d$ counterpart of the
height equation in \cite{munoz:2006} from a physical model of IBS and perform a
systematic numerical analysis of its coarsening properties in relation with
Krug's
conjecture.

Following the standard assumption made in hydrodynamic models of aeolian sand
dunes \cite{dunas}, we will consider that ripples formed under IBS are
translationally
invariant in the $y$ direction; additionally, we assume symmetry under $x
\rightarrow
-x$, as occurs under normal incidence conditions for the
bombarding ions \cite{normal}. As proposed
in \cite{munoz:2006}, the evolution of the
thickness of the mobile surface adatoms layer $R$ and the height of the
bombarded surface $h$ is provided by a pair of coupled equations, namely,
\begin{align}
        \partial_t R &= (1-\phi) \Gamma_{ex} - \Gamma_{ad} + D \partial_x^2 R
\label{eq.R},\\
        \partial_t h &= -\Gamma_{ex}+\Gamma_{ad} \label{eq.h},
\end{align}
where $\Ge$ and $\Ga$ are, respectively, rates of atom excavation from and
addition to the immobile bulk, $(1-\phi)$ measures the fraction of
eroded atoms that become mobile, and the third term in Eq. \eqref{eq.R}
describes thermal diffusion of mobile adatoms.

The rate at which material is sputtered from the bulk is described by
microscopic derivations \cite{teorias} and depends on the local morphology
of the surface
\begin{equation}\label{Ge}
    \Gamma_{ex} = \alpha_0 \big[ 1 + \alpha_{2}
\partial^2_x h
    +\alpha_{3}(\partial_x h)^2 \big],
\end{equation}
where $\alpha_0$ is the sputtering rate for a planar surface.
The rate of nucleation is also related to the local shape of the surface, and is
given by
\begin{equation}\label{Ga}
\Gamma_{ad}=\gamma_0\left[R(1+\gamma_{2}\partial^2_xh)-R_{eq}\right],
\end{equation}
where $R_{eq}$ is the thickness due to the mobile atoms that are
thermally generated even in the absence of bombardment, and
$\gamma_0^{-1}$ is the average time between nucleation events.
After a multiple scale expansion of \eqref{eq.R}-\eqref{Ga}, $R$
can be adiabatically eliminated from Eqs.\ \eqref{eq.R} and
\eqref{eq.h} \cite{castro05, munoz:2006}, obtaining, to lowest
order near threshold of the morphological instability, the
following $1d$ equation for the evolution of the surface height,
\begin{equation}
 \partial_t h(x,t)= - \nu \partial_x^2 h - K \partial_x^4 h  + \lambda_1
 (\partial_x h)^2- \lambda_2 \, \partial_x^2(\partial_x h)^2 \label{eq.},
\end{equation}
where parameters are related to those in \eqref{eq.R}-\eqref{Ga}
and depend on the experimental conditions. To the best of our
knowledge, the {\em deterministic} Eq.\ \eqref{eq.} has not been
systematically studied. We will restrict ourselves to positive
values of $\nu$ and $K$, which are required in order to produce a
long-wavelength instability. Moreover, $\lambda_1$ and $\lambda_2$
are required to have the same sign for mathematical
well-posedness, as shown in \cite{raible:2000b, castro:2005b}. If
the signs of the non-linear terms are simultaneously changed, Eq.\
\eqref{eq.} remains invariant after $h \rightarrow -h$. Thus, we
will only consider positive values of these parameters. For
$\lambda_2=0$, Eq. \eqref{eq.} reduces to the celebrated
Kuramoto-Sivashinsky (KS) equation
\cite{sivashinsky:1983,kuramoto}, which is a paradigm of
spatio-temporal chaos.
Its nonlinear term stabilizes the system and a (disordered)
pattern develops that is characterized by a wavelength that does
{\em not} coarsen, and by chaotic cell dynamics. On large length
scales, the KS system can be effectively described by the {\em
stochastic} Kardar-Parisi-Zhang (KPZ) equation \cite{sneppen92},
paradigmatic of kinetic roughening. In particular, the surface
roughness (global rms width, $W$) \cite{width} for a KPZ interface
scales as a power law with the lateral system size $L$. On the
other hand, for $\lambda_2\neq0$ and $\lambda_1=0$, Eq.\
\eqref{eq.} reduces to the ``conserved" KS equation. This equation
has been studied in the context of amorphous thin film growth
\cite{raible:2000b} and step dynamics on vicinal surfaces
\cite{frisch:2006}; in this case, the linear instability evolves
into an {\em ordered} pattern of paraboloids with {\em
uninterrupted} coarsening.

In order to reduce the number of parameters and simplify the
analysis of \eqref{eq.}, we rescale $x$, $t$ and $h$ by $(K/\nu)^{1/2},
K/\nu^2$ and
$\nu/\lambda_1$, respectively, resulting into a single-parameter equation,
namely,
\begin{equation}
 \partial_t h(t,x)= -\, \partial^2_x h - \partial^4_x h  +
 (\partial_x h)^2- r \, \partial^2_x(\partial_x h)^2 \label{eq.1d},
\end{equation}
where $r=(\nu
\lambda_2)/(K \lambda_1)$ is the (squared) ratio of a linear crossover
lengthscale to a non-linear crossover lengthscale. We have performed a
numerical integration of
\eqref{eq.1d} using a fourth-order Runge-Kutta method and the
improved spatial discretization introduced by Lam and Shin
\cite{lam:1998} for the nonlinear terms. We have used periodic
boundary conditions, lattice constant $\Delta x=0.5$ and time
step $\Delta t=0.01$, checking that results do not differ significantly for
smaller space and time steps. The initial height values were chosen uniformly
distributed between $0$ and $1$ and statistical data were obtained as averages
over 250 random initial conditions. The standard system size of our
simulations was $L=512$ ($1024$ nodes), except when other values are
indicated, and the parameters fixed to $\nu=1$, $K=1$,
$\lambda_1=0.1$, varying $\lambda_2$ in order to check for the
different values of $r$.

\begin{figure}[t]
\begin{center}
\includegraphics[width=0.48\textwidth,clip=]{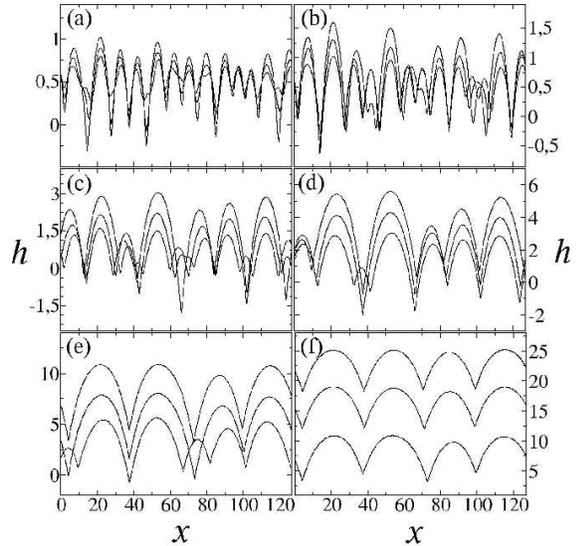}
\end{center}
\caption{Height profiles for Eq.\ \eqref{eq.1d} with $r=50$ on a system
size $L=128$ at times (a) $t=2$, $6$, $11$; (b) $t=11$, $24$, $41$;
(c) $t=41$, $70$, $103$; (d) $t=103$, $153$, $205$; (e) $t=205$,
$302$, $425$; (f) $t=425$, $764$, $1024$. Height profiles at
different times evolve by increasing the maximum value of $h$ with time.
All units are arbitrary.}
\label{fig1}
\end{figure}

\begin{figure}[t]
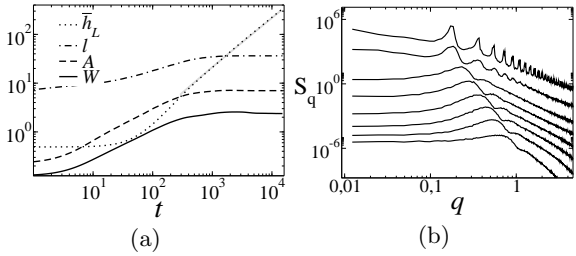

\begin{center}
\begin{minipage}{0.47\linewidth}
\includegraphics[width=\textwidth]{fig2_a.eps}
{(a)}
\end{minipage}
\begin{minipage}{0.47\linewidth}
\includegraphics[width=\textwidth]{fig2_b.eps}
{(b)}
\end{minipage}
\caption{(a) Time evolution of mean surface height average
$\bar{h}_L(t)$, wavelength $l(t)$, amplitude $A(t)$, and global
width $W(t)$, for $r=50$; (b) corresponding surface structure
factor as a function of wave number $q$ at times
$t=4,15,30,60,125,250,1129,$ and $10653$, bottom to top (curves
are offset vertically). All units are arbitrary.} \label{fig2}
\end{center}
\end{figure}

In Fig.\ \ref{fig1} the evolution of the height profile is
depicted for $r=50$. Starting from an initial random distribution,
a periodic surface structure with a wavelength of about the
maximum of the linear dispersion relation \cite{cross:1993},
namely $l_{linear}=2\sqrt{2}\pi$, arises and the amplitude of $h$
increases (Fig.\ \ref{fig1}a). At later stages, the {\em
conserved} KPZ nonlinearity $\partial^2_x(\partial_x h)^2 $ (cKPZ)
induces coarsening of the ordered cell-like structure, wherein the
cells grow in width and height and the number of cells decreases.
This coarsening is such that smaller cells are ``eaten" by larger
neighbors (Figs.\ \ref{fig1}b-e). We show in Fig.\ \ref{fig2}a the
time evolution of the mean height $\bar{h}_L(t)=1/L\sum_x h(x,t)$,
the wavelength $l(t)$, and the amplitude $A(t)$ of the pattern
defined as the mean lateral distance between two consecutive local
minima and the mean vertical distance from a local minimum to the
next local maximum, respectively. As seen in Fig.\ \ref{fig2}a,
for $t \gtrsim 70$, the {\em non-conserved} KPZ term $(\partial_x
h)^2 $ becomes relevant and the mean height of the surface
$\bar{h}_L$ starts increasing to reach a constant velocity (Fig.\
\ref{fig1}f). At the same time, the coarsening process slows down
until stopping completely in the stationary state. This behavior
suggests that the cKPZ term, which acts at small scales, induces
the order and the coarsening process until the slopes and the
characteristic wavelength of the pattern are large enough to make
the KPZ term no longer negligible. At this time, the KPZ term
interrupts the coarsening process, as claimed in \cite{castro06,
munoz:2006}, and a constant average velocity value is achieved as
a consequence of a constant average of slopes across the
interface. Thus, the final wavelength of the pattern depends on
the interplay between the cKPZ and the KPZ terms. Fig.\
\ref{fig2}a also shows the global surface rms width or roughness
$W(t)$ \cite{width}, and Fig.\ \ref{fig2}b shows the surface
structure factor, defined as $ S({q},t)=\left\langle
\hat{h}({q},t) \hat{h}({-q},t) \right\rangle,$ where
$\hat{h}({q},t)$ is the Fourier transform of the field $h(x,t)$.

\begin{figure}[t]
\begin{center}
\begin{minipage}{0.47\linewidth}
\includegraphics[width=\textwidth]{fig3_a.eps}
{(a)}
\end{minipage}
\begin{minipage}{0.47\linewidth}
\includegraphics[width=\textwidth]{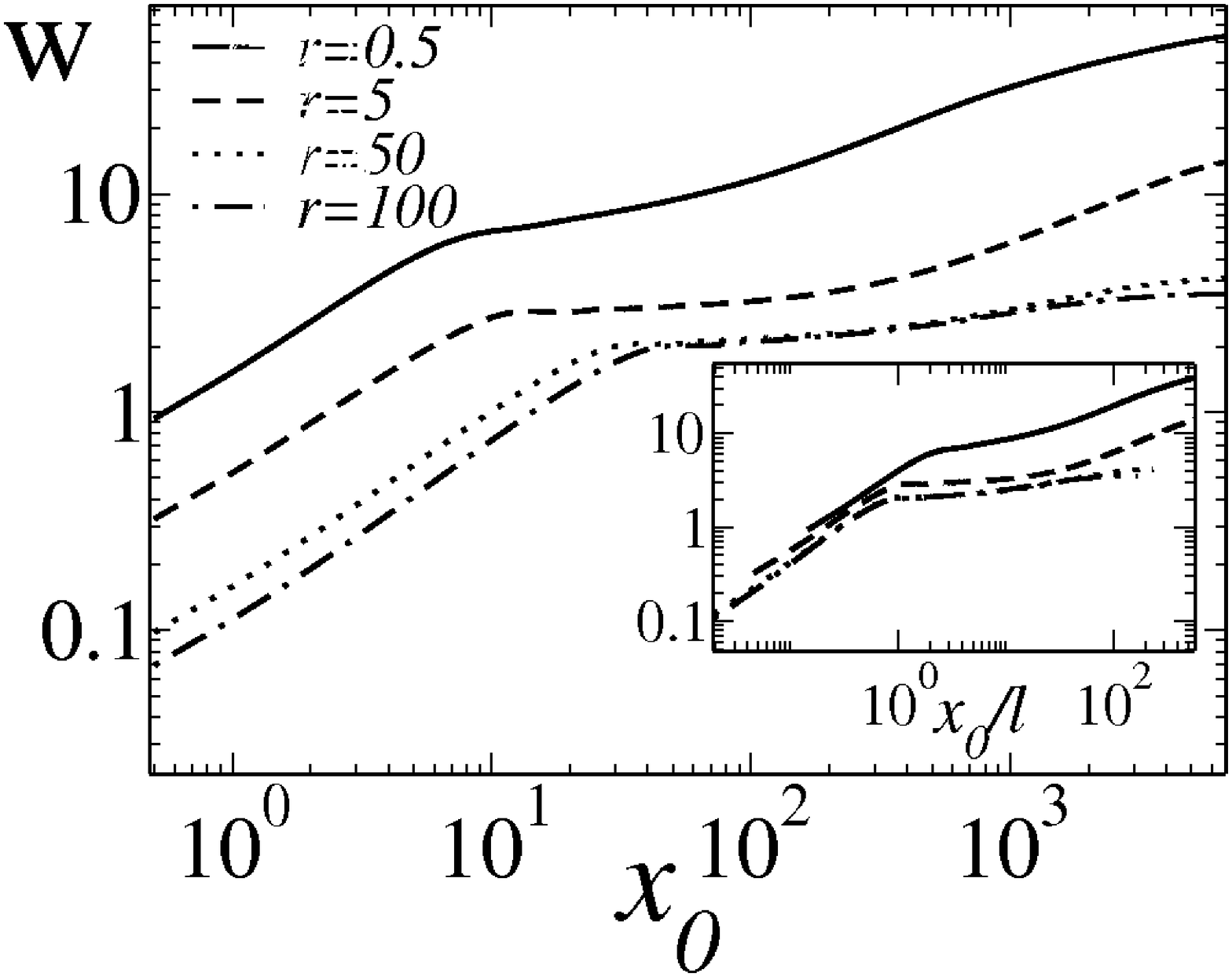}
{(b)}
\end{minipage}
\caption{(a) Height profile at $t=15000$ for a system size
$L=10000$ and $r=50$; (b) local width vs window size at $t=15000$
for several values of $r$ and $L=8192$. Inset: same data plotted
vs $x_0/(2\sqrt{6r})$. All units are arbitrary.} \label{fig3}
\end{center}
\end{figure}

\begin{figure}[t]
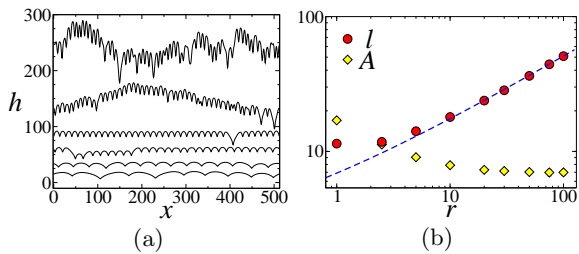

\begin{center}
\begin{minipage}{0.47\linewidth}
\includegraphics[width=\textwidth]{fig4_a.eps}
{(a)}
\end{minipage}
\begin{minipage}{0.47\linewidth}
\includegraphics[width=\textwidth]{fig4_b.eps}
{(b)}
\end{minipage}
\caption{(a) Height profiles at $t=15000$ for (top to bottom) $r=0.1, 0.5, 1,5,
10, 50,$ and $100$; (b) stationary wavelength $l$ and amplitude $A$ as functions
of $r$ (statistical errors are smaller than the symbol sizes). The dashed line
corresponds to $l=2\sqrt{6r}$. All units are arbitrary.}
\label{fig4}
\end{center}
\end{figure}

The first peak in the structure factor indicates the dominant
wavelength of the pattern. We can see how this peak moves to
larger wavelengths (coarsening) until a fixed mode is reached
which corresponds to the stationary value $l \approx 36$ in Fig.\
\ref{fig2}a. At this time coarsening interrupts and the amplitude
saturates to a constant value. Nevertheless, for large enough $L$
and at long distances, the profile disorders in heights (Fig.\
\ref{fig3}a), although the lateral cell-like order is still
preserved for intermediate distances (Fig.\ \ref{fig3}a, inset; note the
difference in scales between the $x$ and $y$ axes). This disorder
reflects in the power law behavior of $S(q,t)$ for $q$ much
smaller than $2\pi/l$ and long enough times (Fig.\ \ref{fig2}b)
or, equivalently, in the behavior of the local width $w(x_0)$
\cite{width} at long times, displayed in Fig.\ \ref{fig3}b as a
function of window size, $x_0$. Due to the parabolic shape of the
cells, the local width scales as $w \sim x_0^2$ for distances
smaller than the cell size, reaching a plateau for intermediate
distances, and finally increasing as a (smaller) power of $x_0$
for large enough distances (kinetic roughening). The plateau in
$w(x_0)$ is related to the lateral order of the pattern, reaching
up to several tens of cells (see e.g. in Fig. \ref{fig4}a an
ordered domain containing over 30 cells for $r=10$). The effective
exponent characterizing the long distance behavior of $w(x_0)$
increases towards its KS value (1/2) for decreasing $r$. Indeed,
the profile is more disordered for small $r$, as can be observed
from Fig.\ \ref{fig4}a, to the extent that for $r \approx 0.2$,
and independently of $L$, the secondary peaks in the structure
factor vanish completely (not shown), and only a weak peak about
the linear instability persists, as in the KS equation. Thus, the
KPZ term is seen to act at larger scales and is responsible for
the disorder of the profile, while the cKPZ terms dominates at
smaller scales with a trend to order the cells vertically.

As indicated above, for large $r$ values and after the initial
times, the cell-like structure is well ordered and the length
scales of the profile are large enough so that the linear fourth
order derivative can be effectively neglected. Thus, we can
rescale $x\rightarrow r^{1/2}x$, $t\rightarrow rt$ and
$h\rightarrow h $ to obtain an effective parameter-free equation.
This means that, for large values of $r$, the solution of Eq.\
\eqref{eq.1d} remains unchanged if we rescale lengths by $r^{1/2}$
and times by $r$. In order to check this hypothesis we present in
Fig.\ \ref{fig4}b the final amplitude and wavelength of the
structure for different values of $r$. We can observe that, for
large values of $r$, the height amplitude does not change with $r$
and the wavelength $l$ is proportional to $r^{1/2}$ as predicted
above. This behavior can be also seen in the inset of Fig.\
\ref{fig3}b, where we rescale the horizontal axis by
$l=2\sqrt{6r}$ (see below), obtaining almost perfect collapse of
the curves for large $r$.

\begin{figure}[t]
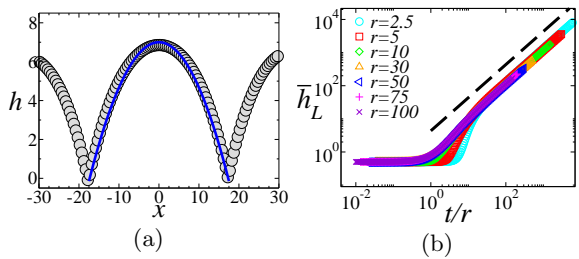

\begin{center}
\begin{minipage}{0.47\linewidth}
\includegraphics[width=\textwidth]{fig5_a.eps}
{(a)}
\end{minipage}
\begin{minipage}{0.47\linewidth}
\includegraphics[width=\textwidth]{fig5_b.eps}
{(b)}
\end{minipage}
\caption{(a) Stationary height profile at $t=15000$ for $r=50$.
The solid line represents the approximate solution given by
\eqref{parabola} for $l=2\sqrt{6r}$ and $ A=7$; (b) mean height
for different values of $r$ as a function of $t/r$. The dashed
line is given by \eqref{hmed}. All units are arbitrary.}
\label{fig5}
\end{center}
\end{figure}

Further progress can be made by recalling that the solution of \eqref{eq.} with
$\lambda_1=0$ is a periodic juxtaposition of parabolas of the form
\cite{raible:2000b,frisch:2006}
\begin{equation}
h(x)=A-\frac{4A}{l^2}\,x^2. \label{parabola}
\end{equation}
We can consider this function as an approximate solution of Eq.\
\eqref{eq.1d} for large values of $r$. In Fig.\ \ref{fig5}a we
compare the numerical profile with the function \eqref{parabola}
for $r=50$. Using this solution and assuming as a condition for
the final (interrupted) structure that the averaged nonlinear
contributions along one period must be equal, we obtain a relation
between the final wavelength and $r$, namely $l=2\sqrt{6r}$. This
function is represented in Fig.\ \ref{fig4}b, fitting accurately
the stationary wavelengths obtained numerically for large values
of $r$. The net mean growth velocity of the height average is only
due to the KPZ nonlinear term and is given by $ v= \left \langle
\lambda_1 \overline{[\partial_x h(x)]^2}_L \right \rangle$.\
Assuming \eqref{parabola} as an approximate solution and
integrating over one period, we obtain the net mean profile
evolution. It reads
\begin{equation}
\bar{h}_L(t)=vt\approx\lambda_1 \left\{ \frac{1}{l}
\int^{l/2}_{-l/2} \left[\partial_x h(x)\right]^2  dx  \right\} t=
\lambda_1 \frac{11}{r}t, \label{hmed}
\end{equation}
where we have substituted $l=2\sqrt{6r}$, and $A\approx 7$ is
obtained from Fig.\ \ref{fig4}b. As seen from Fig.\ \ref{fig5}b,
for large values of $r$ the evolution of $\bar{h}_L$ becomes
$r$-independent if we rescale $t\rightarrow rt$, as already
indicated. Furthermore, at long times, the growth velocity given
by \eqref{hmed} agrees accurately with the numerical observations
for a wide range in $r$.

In conclusion, we have derived a novel deterministic $1d$ equation
with shift symmetry in the context of ion-beam sputtering and
performed a numerical analysis to obtain some relevant information
about its solutions. For large values of $r$ we have estimated the
final pattern wavelength assuming a parabolic solution analogous
to that of the ``conserved" KS equation and checked our assumption
by comparing with the numerical mean height evolution. The
resulting single-parameter Eq.\ \eqref{eq.1d} interpolates between
the KS equation for $r=0$, which presents a chaotic solution and
no coarsening, and the ``conserved" KS equation for $r\rightarrow
\infty$ (upon rescaling $h\rightarrow h/r$), which displays
unbounded coarsening. This behavior is similar to the $1d$ {\em
convective} Cahn-Hilliard (cCH) equation studied in
\cite{golovin01}. However, as reported in Ref.\ \cite{golovin01}, coarsening
does not interrupt in the cCH system for a whole range of parameter values but, 
rather, proceeds logarithmically for long times, whereas Eq.\ 
\eqref{eq.1d} does present interrupted coarsening with an {\em ordered 
pattern} of constant wavelength and amplitude {\em for intermediate distances}.
This behavior provides a new scenario for the classification of surface
coarsening phenomena in Ref.\ \cite{politi}, where all the studied evolution
equations display perpetual coarsening, or else develop a pattern with a frozen
wavelength while the amplitude continues growing without bound in
the course of time. Regarding Krug's conjecture \cite{krug:2001},
the present example reinforces its validity for long range order
properties, since the pattern produced by Eq.\ \eqref{eq.1d}
disorders at large length scales. However, our results suggest
that it is still possible to stabilize a well ordered pattern over
(intermediate) distances ranging even up to several tens of cell
sizes in contrast with other equations in this context.

\begin{acknowledgments}
This work has been partially supported by MEC (Spain), through
Grants Nos.\ BFM2003-07749-C05, -01, -05, and the FPU programme
(J.M.-G.).

\end{acknowledgments}

\bibliography{bib}

\begin{thebibliography}{23}
\bibitem[{not({\natexlab{a}})}]{note1}
\bibinfo{note}{We take ``interface" or ``height" equation in the restricted
  sense that physics is invariant under global height shifts $h(x,t)\rightarrow
  h(x,t)+const$.}
\end{thebibliography}

\end{document}